\documentclass[journal]{IEEEtran}

\ifCLASSINFOpdf
   \usepackage[pdftex]{graphicx}
\else
\fi

\ifCLASSOPTIONcompsoc
 \usepackage[caption=false,font=normalsize,labelfont=sf,textfont=sf]{subfig}
\else
 \usepackage[caption=false,font=footnotesize]{subfig}
\fi

\usepackage{amssymb}
\usepackage{amsmath,amssymb,amsfonts}
\usepackage[nolist]{acronym}
\usepackage{steinmetz} 
\usepackage{cite}
\usepackage{multirow}
\usepackage{enumitem} 
\setlist[description]{
	leftmargin=30pt, 		
	topsep=0pt,             
	itemsep=0pt,            
	labelindent=0pt,
	style=multiline,
}

\usepackage{algorithm} 
\usepackage{algorithmicx}
\usepackage{algpseudocode}

\usepackage[usenames,dvipsnames]{xcolor}

\acused{LP-DOE}
\acused{NLP-DOE}

\begin{acronym}
	\acro{13NF}{IEEE 13 Node Test Feeder}
	\acro{123NF}{IEEE 123 Node Test Feeder}
	\acro{ACT}{Australian Capital Territory}
	\acro{AC}{alternating current}
	\acro{ANU}{Australian National University}
	\acro{CIM}{current injection method}
	\acro{DER}{distributed energy resource}
	\acrodefplural{DER}[DERs]{distributed energy resources}
    \acro{DOE}{dynamic operating envelope}
	\acrodefplural{DOE}[DOEs]{dynamic operating envelopes}
    \acro{DSO}{distribution system operator}
	\acrodefplural{DSO}[DSOs]{distribution system operators}
    \acro{EV}{electric vehicle}
	\acrodefplural{EV}[EVs]{electric vehicles}
	\acro{ESS}{energy storage system}
	\acrodefplural{ESS}[ESSs]{energy storage systems}
	\acro{IEC}{International Electrotechnical Commission}
    \acro{LACE}{Linear Analytical Calculation of Envelopes}
    \acro{LP}{linear programming}
    \acro{LP-DOE}{LP-DOE}
	\acro{L-QP}{Local-Quadratic Program}
	\acro{LV}{low-voltage}
	\acro{MV}{medium-voltage}
    \acro{MPC}{model predictive control}
    \acro{NLP}{non-linear programming}
    \acro{NLP-DOE}{NLP-DOE}
	\acro{OPF}{optimal power flow}
	\acro{OPF-PC}{OPF-based Proportional Control}
	\acro{PCC}{point of common coupling}
	\acro{PF}{power flow}
	\acro{PV}{photovoltaic}
	\acro{QP}{quadratic program}
	\acro{QCQP}{quadratically constrained quadratic program}
    \acro{RHO}{receding-horizon optimization}
	\acro{SOC}{state of charge}
	\acro{TOU}{time-of-use}
	\acro{VB-VVW}{Voltage-Based Volt-Var-Watt}
\end{acronym}

\begin{document}
%

\title{Allocation of Dynamic Operating Envelopes \\ in Radial Distribution Networks}

%
%

\author{Wilhiam~de~Carvalho,
        Florin~Capitanescu, 
        Cyril~Rasic, 
        Jean-François~Toubeau, 
        François~Vallée
\thanks{W. de Carvalho and F. Capitanescu are with the Luxembourg Institute of Science and Technology (LIST), Luxembourg (e-mail: wilhiam.carvalho, florin.capitanescu @list.lu). C.~Rasic, J.~Toubeau and F.~Vallee are with the University of Mons, Belgium.}
\thanks{This paper is a preprint of a paper accepted by CIRED 2026 and is subject to Institution of Engineering and Technology Copyright. When the final version is published, the copy of record will be available at IET Digital Library.}
}%

\markboth{}%
{}

\IEEEaftertitletext{\vspace{-13pt}}

\maketitle

\begin{abstract}

This paper provides an in-depth analysis on how different aspects of the dynamic operating envelope (DOE) formulation impact the computation and allocation of network capacity. We show that the envelopes are significantly affected by the power flow model (non-linear or linear), binding network constraint (thermal or voltage) and by the calculation case (import or export envelope). We also propose a novel DOE algorithm (LACE) that presents transparent and scalable computation that is useful for larger networks or to act in tandem with other optimization engines. We run numerical simulations with different test feeders, including a realistic low-voltage feeder with real-world data from Belgium. This paper provides crucial insights and tools to distribution system operators (DSOs), stakeholders and academics alike to make sure DOE calculation achieves desirable and efficient outcome.

\end{abstract}


\section{Introduction}

The accelerated adoption of \acp{DER} is challenging the safe and efficient operation of distribution networks. Traditional grid management and planning tools no longer capture the fast-changing behavior of modern networks, often resulting in over-conservative restrictions. In contrast, the recently proposed \acp{DOE} \cite{petrou2021} accounts for variable grid conditions and dynamically adjusts operating limits of \acp{DER} to maintain a reliable network operation. \acp{DOE} have shown great promise in unlocking non-utilized network capacity and allowing a more efficient and secure operation of distribution systems \cite{michliu2022}.

\ac{DOE} computation is based on optimization engines that find the maximum available grid capacity at a given time that can be allocated in the form of envelopes (power limits) to nodes across the network (see Fig.~\ref{fig:DOE_overview}). However, different optimization formulations and network aspects drastically change \ac{DOE} calculation and allocation. Understanding how envelopes are allocated to different locations on the network is challenging and crucial to make sure appropriate amount is assigned to prosumers with readily available controllable power, preventing over- and under-allocation of capacity.

Existing works have explored the impact of \ac{DOE} objective functions on the allocation of envelopes \cite{petrou2021}, \cite{edge2023reportfair}, \cite{moring2024}, but other crucial aspects of the optimization problem and the analytical understanding are yet to be unveiled. Previous research emphasizes that upstream nodes closer to the substation have technical advantages \cite{petrou2021}, \cite{michliu2022}, \cite{moring2024}, without accounting for fundamental differences in optimization formulations that may result in completely distinct allocation outcomes.

In this paper, we provide an in-depth analysis on how different aspects of the optimization formulation impact the calculation and allocation of envelopes. We show that \acp{DOE} are significantly affected by the power flow model (non-linear or linear), binding network constraint (thermal or voltage) and by the calculation case (import or export envelope). We show that \ac{DOE} calculation presents counter-intuitive allocation outcomes due to the non-linear or linear behavior of the chosen model. We also propose a novel analytical algorithm to obtain \acp{DOE} that present simpler computation and higher performance. We provide mathematical derivations and numerical simulations on several test feeders, including a realistic \ac{LV} feeder with an anonymized dataset from Belgium.

\section{Optimization-based DOE Calculation}

\begin{figure}[!t]
    \centering
	\includegraphics[width=0.45\textwidth]{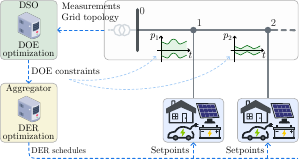}
	\caption{Calculation and allocation of \acp{DOE} in a distribution network.}
	\vspace{-0pt}
	\label{fig:DOE_overview}
\end{figure}

As in Fig.~\ref{fig:DOE_overview}, let node~0 represent the substation (slack) bus, $\mathcal{N}:=\{1,2,...,N\}$ denote the set of other nodes, and $\mathcal{E}$ the set of all line segments. For each line segment $(l,m) \in \mathcal{E}$, let $P_{lm}$ denote the real power, $Q_{lm}$ the reactive power and $l_{lm}$ the current magnitude squared. For each node $m \in \mathcal{N}$, load is represented by the base load $\widetilde{p}_{m} +j \widetilde{q}_{m}$ (known, non-controllable) plus the flexible load $p_m$ (controllable), where $p_m$ is the decision variable of interest that represents the envelope bounds.

We adopt the branch flow equations \cite{moring2024}, \cite{baran1989sizing} widely used to model distribution grids. The full AC power flow equations for every line segment $(l,m) \in \mathcal{E}$ are 
\begin{equation}
P_{lm} = r_{lm} l_{lm} + p_{m} + \widetilde{p}_{m} + \sum\nolimits_{\scriptscriptstyle n:(m,n) \in \mathcal{E}} P_{mn},
\label{eq:distflow_p}
\end{equation}
\begin{equation}
Q_{lm} = x_{lm} l_{lm} + \widetilde{q}_{m} + \sum\nolimits_{\scriptscriptstyle n:(m,n) \in \mathcal{E}} Q_{mn},
\label{eq:distflow_q}
\end{equation}
\begin{equation}
P^2_{lm} + Q^2_{lm} = U_{l} l_{lm},
\label{eq:distflow_pq}
\end{equation}
\begin{equation}
U_{l} = U_{m} + 2r_{lm} P_{lm} + 2x_{lm} Q_{lm} - (r_{lm}^2 + x_{lm}^2)  l_{lm},
\label{eq:distflow_v}
\end{equation}
where $U_m = V_m^2$ is the voltage magnitude squared. Model \eqref{eq:distflow_p}-\eqref{eq:distflow_v} results in non-convex optimization problems and this often motivates the formulation of linear equations to support convex optimization problems \cite{petrou2021}, \cite{michliu2022}, \cite{baran1989sizing}. The linear model disregard the power loss in line segments, simplifying to
\begin{equation}
P_{lm} = p_{m} + \widetilde{p}_{m} + \sum\nolimits_{\scriptscriptstyle n:(m,n) \in \mathcal{E}} P_{mn},
\label{eq:lindistflow_p}
\end{equation}
\begin{equation}
Q_{lm} = \widetilde{q}_{m} + \sum\nolimits_{\scriptscriptstyle n:(m,n) \in \mathcal{E}} Q_{mn},
\label{eq:lindistflow_q}
\end{equation}
\begin{equation}
U_{l} = U_{m} + 2r_{lm} P_{lm} + 2x_{lm} Q_{lm}.
\label{eq:lindistflow_v}
\end{equation}

\ac{DOE} calculation is divided in two cases: the import and the export case. The import envelope represents the additional consumption ($p_m \geq 0$) that nodes are allowed, whereas the export case represent the additional injection (generation) ($p_m \leq 0$) that nodes are allowed to before breaching network limits. The import \ac{DOE} is obtained by solving
\begin{subequations}
\begin{align}
\max_{\substack{p_m}} \quad  & \sum\nolimits_{\scriptscriptstyle m \in \mathcal{N}} p_{m} \\
\textrm{s.t.}
\quad & \eqref{eq:distflow_p}-\eqref{eq:distflow_v} \mbox{ or }  \eqref{eq:lindistflow_p}-\eqref{eq:lindistflow_v}, \\
\quad & P_{01}^2 + Q_{01}^2 \leq \overline{S}_{01}^2, \\
\quad & \underline{U} \leq U_{m} \leq \overline{U}, & \forall m \in \mathcal{N}, \\
\quad & \underline{p}_m \leq p_m \leq \overline{p}_m, & \forall m \in \mathcal{N}, \\
\quad & p_m \geq 0, & \forall m \in \mathcal{N}.
\end{align}
\label{eq:conv_doe_imp}
\end{subequations}
\hspace{-5pt} Problem \eqref{eq:conv_doe_imp} maximizes the combined import envelope across the network. Constraint (\ref{eq:conv_doe_imp}b) represents the power flow model, (\ref{eq:conv_doe_imp}c) is the thermal limit at the feeder head, (\ref{eq:conv_doe_imp}d) is the voltage constraint, (\ref{eq:conv_doe_imp}e) is the envelope physical bounds, and (\ref{eq:conv_doe_imp}f) allows only consumption (import envelope case). The export \ac{DOE} is calculated by simply changing the objective to a minimization and (\ref{eq:conv_doe_imp}f) to $p_m \leq 0$ \cite{moring2024}. 

Although time index $(t)$ is not explicitly shown, \acp{DOE} are computed every time period (e.g. 15-min) with up-to-date measurements and forecasts of known variables such as $V_0, \widetilde{p}_n, \widetilde{q}_n$.

Adopting power flow model \eqref{eq:distflow_p}-\eqref{eq:distflow_v} results in a non-convex \ac{NLP}, termed here the \ac{NLP-DOE} approach, whereas \eqref{eq:lindistflow_p}-\eqref{eq:lindistflow_v} results in a convex \ac{LP}, termed here \ac{LP-DOE} approach. In any case, considering that (\ref{eq:conv_doe_imp}e) is large enough, the solution is given when all spare grid capacity runs out and network constraints are binding. As such, either thermal or voltage (or both) constraints are active at the optimal solution. The optimal solution favors nodes that have locational advantages that allows a larger \textit{combined} consumption/injection ($p_m$).

\section{DOE Allocation Analysis}

We re-formulate problem \eqref{eq:conv_doe_imp} and show how thermal and voltage constraints individually impact the calculation and allocation \acp{DOE}. Theoretical analysis is provided for the linear model, while empirical results are used for the non-linear case. Starting first with \ac{LP-DOE}, we define voltage deviation by $E_m := U_0 - U_m$ and write \eqref{eq:lindistflow_p}-\eqref{eq:lindistflow_v} in the compact, matrix format similar to \cite{decarvalho2024arxiv} as
\begin{align}
\boldsymbol{E} = \boldsymbol{R} \boldsymbol{p} + \boldsymbol{\widetilde{E}} ,
\label{eq:lin_compact2}
\end{align}

\noindent where $\boldsymbol{\widetilde{E}} = \boldsymbol{R} \boldsymbol{\widetilde{p}} + \boldsymbol{X} \boldsymbol{\widetilde{q}}$ is the underlying voltage deviation associated with the uncontrollable (known) load. Column vectors $\boldsymbol{E}$, $\boldsymbol{p}$, $\boldsymbol{\widetilde{p}}$, $\boldsymbol{\widetilde{q}} \in \mathbb{R}^{\text{N}}$ collects the variables of each node $m \in \mathcal{N}$ (e.g., $\boldsymbol{p} = [p_1, p_2, ..., p_N]^\top$). Matrices $\boldsymbol{R}, \boldsymbol{X}$ represents the common path impedance between nodes as shown in \cite{decarvalho2024arxiv}.

Essentially, \eqref{eq:lin_compact2} represents the grid voltage deviation as a function of the real power envelope $\boldsymbol{p}$. We define $\overline{E}:= U_0 - \underline{U}$ as the upper bound and $\underline{E}:=U_0 - \overline{U}$ as the lower bound on voltage deviation. Let the non-controllable real power at the feeder head be $\widetilde{P}_{01} = \sum_{\scriptscriptstyle m \in \mathcal{N}} \widetilde{p}_{m}$ and the reactive power be $\widetilde{Q}_{01} = \sum_{\scriptscriptstyle m \in \mathcal{N}} \widetilde{q}_{m}$. We represent the thermal limit by $\overline{P}_{01} = \sqrt{ \overline{S}_{01}^2 - \widetilde{Q}_{01}^2 }$, where $\overline{S}_{01}$ is the nominal apparent power of the substation transformer.

\subsection{\ac{LP-DOE}: Thermal constraint as the limiting factor}

We define the \textit{spare import thermal capacity} at the feeder head by $\widehat{P}_{01} := \overline{P}_{01} - \widetilde{P}_{01} \geq 0$, i.e., the difference between maximum allowed and the underlying power. To fully capture how the network constraints impact calculation and allocation of \acp{DOE}, we consider that constraint (\ref{eq:conv_doe_imp}e) is not active and that every node can receive envelopes. As such, the import \ac{DOE} calculation for a thermally-constrained case is obtained by reformulating \eqref{eq:conv_doe_imp} as
\begin{equation}
\max \{ \boldsymbol{1}^\top \boldsymbol{p} \mid \boldsymbol{1}^\top \boldsymbol{p} \leq \widehat{P}_{01}, \, \boldsymbol{p} \geq \mathbf{0} \} .
\label{eq:import_doe_therm_const}
\end{equation}
The solution of \eqref{eq:import_doe_therm_const} is given by the set of optimizers $\{ \boldsymbol{1}^\top\boldsymbol{p}^* = \widehat{P}_{01} : \boldsymbol{p}^* \geq \mathbf{0} \}$. In more detail, for any network where $N>1$, the \ac{LP} problem is degenerate and it contains infinitely many solutions. In terms of envelope allocation, the \ac{LP-DOE} engine sees no network locational penalties and any fairness (reallocation) strategy results in the same combined import value.

We define the \textit{spare export thermal capacity} at the feeder head as $\breve{P}_{01} := - \overline{P}_{01} - \widetilde{P}_{01} \leq 0$ and reformulate the export optimization problem similar to \eqref{eq:import_doe_therm_const} to reach the same conclusion. In summary, both the import and the export cases have no network locational penalties when the thermal limit is the binding constraint with the \ac{LP-DOE} approach. In these cases, equally spreading envelopes across the network does not affect the combined amount of envelope. Table~\ref{tab:allocation_outcome} summarizes the envelope allocation outcome.

\subsection{\ac{LP-DOE}: Voltage constraint as the limiting factor}

We define the \textit{spare voltage drop} by $\widehat{E}_m := \overline{E} - \widetilde{E}_m$ and reformulate the import optimization problem \eqref{eq:conv_doe_imp} for the voltage-constrained case as 
\begin{equation}
\max \{ \boldsymbol{1}^\top \boldsymbol{p} \mid \boldsymbol{R} \boldsymbol{p} \leq \boldsymbol{\widehat{E}}, \, \boldsymbol{p} \geq \mathbf{0} \} .
\label{eq:import_vconstrained}
\end{equation}
In radial distribution networks, with node~1 as the only node directly connected to the slack, it holds that the first-row (and column) entries of matrix $\boldsymbol{R}$ is no larger than the entries of any other row of the same column, i.e., $R_{1n} = R_{n1} = 2 r_{01} \leq R_{mn}$ for every $m,n \in \mathcal{N}$. Thus we can demonstrate that $\boldsymbol{p}^* = (B, 0, ..., 0)$ with $B := \min_m \widehat{E}_m/2r_{01}$ is an optimal and feasible solution of \eqref{eq:import_vconstrained}, but it need not be the unique solution. We introduce 
\begin{align}
    p_n^{solo} := \min\nolimits_{\scriptscriptstyle m \in \mathcal{N}} \big( \widehat{E}_m / R_{mn} \big),
    \label{eq:psolo_import}
\end{align}
representing the import envelope for node $n \in \mathcal{N}$ when all other envelopes are zero. That is, $p_n^{solo}$ gets the tightest bound of the voltage constraint when all other $p_k=0, \ k \neq n$. If there are two or more nodes with the largest $p_n^{solo}$, then the solution is not unique and multiple upstream nodes have the locational preference.

\begin{table}[!t]
\footnotesize
\renewcommand{\arraystretch}{1.1}
\caption{Summary of DOE Allocation Outcome.}
\label{tab:allocation_outcome}
\begin{tabular}{|c|c|c|c|}
		\hline
		\textbf{DOE case}                     &\textbf{Binding constr.}           & \textbf{\ac{LP-DOE}}      & \textbf{\ac{NLP-DOE}}     \\ \hline
		\multirow{2}{*}{\textbf{Import}}      &\textbf{Thermal}                   & Any node                  & Upstream nodes            \\ \cline{2-4}
	    		                                 &\textbf{Under-voltage}             & Upstream nodes            & Upstream nodes            \\ \hline
        \multirow{2}{*}{\textbf{Export}}      &\textbf{Thermal}                   & Any node                  & Downstream nodes          \\ \cline{2-4}
            		                           &\textbf{Over-voltage}              & Upstream nodes            & Upstream nodes            \\ \hline
	\end{tabular}
\end{table}

Similarly for the export envelope case, we define the \textit{spare voltage rise} as $\breve{E}_m := \underline{E} - \widetilde{E}_m$ and formulate a problem similar to \eqref{eq:import_vconstrained} and \eqref{eq:psolo_import}. In summary, node~1 is always an optimal solution, but it can be shared between other closer nodes that are equally optimal, as determined by \eqref{eq:psolo_import}.

It is also possible that both thermal and voltage constraints are active at the same time. In this case, nodes upstream also have the locational advantage, since thermal constraints add no locational penalty that competes with voltage-constrained penalties.

\subsection{\ac{NLP-DOE}: Thermal constraint as the limiting factor}

Achieving general mathematical conclusions with the \ac{NLP-DOE} is non-trivial as equations are non-linear and non-convex. In this section, we analyze the allocation of envelopes based on empirical analysis. The \ac{NLP-DOE} includes the non-linear terms corresponding to line losses of real and reactive power. In general, line losses penalize the nodes that results in further voltage deviation or larger power flow at the feeder head.

\begin{figure}[!t]
	\centering
	\subfloat[]{%
		\includegraphics[width=0.45\linewidth]{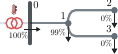}}
	\hfill
	\subfloat[]{%
		\includegraphics[width=0.45\linewidth]{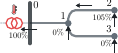}}
	\caption{Allocation of envelopes with the \ac{NLP-DOE} approach for a thermally-constrained feeder: (a) import and (b) export envelope case.}
	\label{fig:nlp_thermal}
\end{figure}

Fig.~\ref{fig:nlp_thermal} illustrates a thermally-constrained case where the spare capacity $\widehat{S}_{01} = 1$~p.u. (100\%) is allocated into envelopes plus a portion of real and reactive line losses. Fig.~\ref{fig:nlp_thermal}(a) depicts a typical import case where the spare thermal capacity is fully allocated to the first node plus 1\% of line losses in the first line segment. In this case, the import envelope is maximized when allocated to nodes that result in smaller line losses (transportation costs). That is often attained by allocating envelopes to upstream nodes with shorter electrical distance.

Fig.~\ref{fig:nlp_thermal}(b) depicts a typical export case with full envelope allocation to node~2, allowing the export power to increase beyond the thermal capacity since part of the injected power is consumed by line losses. Power losses caused by reverse power flows alleviate the stress at the feeder head --- hence the export envelope is maximized when power losses are larger to allow greater injection by nodes. The same optimal result is attained by fully allocating the envelope to node 3, if it has the same base load and the same path impedance from the slack bus as node~2. However, equally splitting the envelope between node 2 and 3 results in smaller combined envelope (52\% at node~1 and 52\% at node~2) as the losses, which increases quadratically with the current magnitude, would also be smaller.

\subsection{\ac{NLP-DOE}: Voltage constraint as the limiting factor}

Similar to \ac{LP-DOE}, nodes with larger path impedance shows higher voltage sensitivity to nodal power consumption/injection. This penalizes the allocation of envelope to nodes further downstream as they present considerably higher electrical distance.

Table~\ref{tab:allocation_outcome} summarizes the envelope allocation outcome. Observe that in voltage-constrained cases, upstream nodes have the allocation preference regardless of power flow model and calculation case (import or export \ac{DOE}). However, thermally-constrained cases present a diverse allocation outcome and locational advantages can completely change depending on the model and calculation case. Designing a fairness policy based on weights to favor downstream nodes, for example, might actually make the fairness issue more acute if the network is often thermally constrained.

\section{LACE Algorithm}

Optimization-based approaches typically rely on external solvers and system integration that can introduce security, computational and reliability challenges in real-world industrial applications. We develop a novel analytical algorithm to directly compute \acp{DOE} based on the linear model, termed \ac{LACE}. Algorithm~\ref{alg:one} summarizes the main calculation steps for the import \ac{DOE} case, and a similar algorithm is used for the export case.

In essence, \ac{LACE} finds the least penalized node based on \eqref{eq:psolo_import} and allocates the most conservative value between three options: (i) envelope bounds; (ii) thermally-constrained capacity; and (iii) voltage-constrained capacity. The envelope bound is provided by the participating user or aggregator as a static limit on the envelope. The thermal and voltage-constrained values are updated every iteration $i$ and they converge to zero as envelopes are allocated.

\ac{LACE} engine has shown to achieve similar results as the \ac{LP-DOE} with shorter computation time and less prone to failures. Since \ac{LACE} does not rely on external 3rd party software or solvers, the analytical calculation can prove to be a reliable and affordable addition to the \ac{DSO} toolbox. We also stress that it is straightforward to include fairness policies into Algorithm~\ref{alg:one}.

\begin{algorithm}[!t]
\caption{LACE — Import Case}
\label{alg:one}
\begin{algorithmic}[1]
\State Initialize iteration counter $i \gets 0$
\State Initialize set of remaining nodes $\mathcal{A}(i)$
\State Compute initial spare thermal capacity $\widehat{P}_{01}(i)$
\State Compute initial spare voltage capacity $\widehat{\boldsymbol{E}}(i)$
\While{$|\mathcal{A}(i)|>0$ and $\widehat{P}_{01}(i)>0$ and $\min_n\widehat{E}_n(i)>0$}
\State Compute $p^{solo}_n$ as in \eqref{eq:psolo_import} for all $n \in \mathcal{A}(i)$
\State Get node number $m$ with largest $p^{solo}$
\State Allocate to node $m$: $p_m \gets \min(\overline{p}_m, \widehat{P}_{01}(i), p^{solo}_m)$
\State Subtract node $m$ from set: $\mathcal{A}(i+1) \gets \mathcal{A}(i) \backslash \{m\} $
\State Update thermal capacity: $\widehat{P}_{01}(i+1) \gets \widehat{P}_{01}(i) - p_m$
\State Update voltage capacity: $\widehat{\boldsymbol{E}}(i+1) \gets \widehat{\boldsymbol{E}}(i) - \boldsymbol{R}_{:,m} p_m$
\State Increment counter: $i \gets i + 1$
\EndWhile
\end{algorithmic}
\end{algorithm}

\section{Numerical Simulations}

We run simulations with a conceptual 3-node and a realistic 8-node \ac{LV} feeder based on a Belgian residential network \cite{mediwaththe2021} with real-world load from the Flobecq dataset \cite{klonari2014}. We also run a scalability experiment by varying the number of nodes of a distribution feeder. Nominal phase-neutral voltage is $0.23$~kV and $V_0$ is kept at 1~p.u. Voltage bounds are set to 0.90 and 1.10~p.u. and we illustrate cases with different thermal capacity values at the feeder head.

Both import and export \acp{DOE} are computed and we compare all three calculation approaches: \ac{LACE} engine described in Algorithm~\ref{alg:one}; \ac{LP-DOE} by solving \eqref{eq:conv_doe_imp} with the linear model \eqref{eq:lindistflow_p}-\eqref{eq:lindistflow_v}; and \ac{NLP-DOE} by solving \eqref{eq:conv_doe_imp} with the full AC non-linear power flow model \eqref{eq:distflow_p}-\eqref{eq:distflow_v}.

We have implemented all three approaches from scratch in Julia/JuMP programming language. GLPK is used for solving \ac{LP-DOE} and Ipopt for solving \ac{NLP-DOE}. We validated our results against the established OpenDSS software. The simulations are run in a HP Elitebook 860 G9 laptop, with a 12th Gen Intel i7-1280P, 1800 Mhz and 32~GB of installed RAM.

\subsection{Conceptual Example: 3-node test feeder}

Fig.~\ref{fig:DOE_overview} depicts the 3-node conceptual feeder. To highlight the allocation behavior of \ac{DOE} engines, we set $r_{01}=r_{12}=0.1 \Omega$ and $x_{01}=x_{12}=0.05 \Omega$. Base load at the nodes are $\widetilde{p}_1 = \widetilde{p}_2 = 4.8$~kW and $\widetilde{q}_1 = \widetilde{q}_2 = 2$~kvar. The nominal power of the transformer is 20~kVA, resulting in a spare import thermal capacity of $\widehat{P}_{01}=10$~kW and spare export of $\breve{P}_{01}=-29.2$~kW for the linear approaches. Each node has a participating prosumer with controllable \acp{DER}.

Table~\ref{tab:doe_3n_thermal} summarizes the numerical results, displayed as export / import, for a thermally-constrained case. Table~\ref{tab:doe_3n_thermal} shows the optimal envelope, as well as the highest and the lowest voltage magnitude on the network and the apparent power flow at the feeder head. Note that the power at the feeder head is at the boundary for both the export and import case, while there is still significant spare voltage capacity. The linear models (\ac{LACE} and \ac{LP-DOE}) resulted in different allocation outcome, but the same optimal combined envelope of -29.2 and 10~kW for export and import, respectively. Any allocation combination between node~1 and 2 with this combined amount is the optimal result for the linear models. 

The \ac{LACE} algorithm starts from nodes with a larger voltage capacity first (see \eqref{eq:psolo_import} and Algorithm~\ref{alg:one}). As for the \ac{LP-DOE}, we have observed that allocation completely changes when switching between the available solver methods (Simplex, Interior Point and Exact). For instance, the GLPK with the Simplex solver resulted in the exactly same numbers as the \ac{LACE} algorithm. In contrast, the \ac{NLP-DOE} resulted in full import allocation to node~1, and full export allocation to node~2, as summarized in Table~\ref{tab:allocation_outcome}. For any other combination of allocation, the combined import and export envelope is significantly reduced (in absolute terms), suggesting that the optimal \ac{NLP} solution is unique. 

Table~\ref{tab:doe_3n_voltage} presents \ac{DOE} results for the voltage-constrained case. To illustrate the voltage-constrained case, the transformer at the feeder head is replaced with a larger capacity of 100~kVA.

\begin{table}[!t]
	\renewcommand{\arraystretch}{1.1}
	\caption{DOE results for thermally-constrained network.}
	\vspace{-0pt}
	\label{tab:doe_3n_thermal}
    \footnotesize
	\begin{tabular}{|c|c|c|c|c|c|}
		\hline
		\textbf{Approach}     & $p_1$ (kW)    & $p_2$ (kW)    & Voltage (p.u.)   & $S_{01}$ (kVA)  \\ \hline
		\textbf{LACE}         & -29.2 / 10    & 0 / 0         & 1.033 / 0.947    & -20 / 20         \\ \hline
	    \textbf{\ac{LP}}      & -20.8 / 3.8   & -8.4 / 6.2    & 1.038 / 0.934    & -20 / 20         \\ \hline
        \textbf{\ac{NLP}}     & 0 / 9.1       & -30.8 / 0     & 1.076 / 0.947    & -20 / 20         \\ \hline
	\end{tabular}
\end{table}

\begin{table}[!t]
	\renewcommand{\arraystretch}{1.1}
	\caption{DOE results for voltage-constrained network.}
	\vspace{-0pt}
	\label{tab:doe_3n_voltage}
    \footnotesize
	\begin{tabular}{|c|c|c|c|c|c|}
		\hline
		\textbf{Approach}     & $p_1$ (kW)    & $p_2$ (kW)    & Voltage (p.u.)   & $S_{01}$ (kVA)  \\ \hline
		\textbf{LACE}         & -67.1 / 32.8  & 0 / 0         & 1.100 / 0.900    & -57.7 / 42.6    \\ \hline
	    \textbf{\ac{LP}}      & -67.1 / 32.8  & 0 / 0         & 1.100 / 0.900    & -57.7 / 42.6    \\ \hline
        \textbf{\ac{NLP}}     & -70.9 / 30.4  & 0 / 0         & 1.100 / 0.900    & -55.8 / 44.2    \\ \hline
	\end{tabular}
\end{table}

\begin{figure}[!t]
	\centering
	\includegraphics[width=0.99\linewidth]{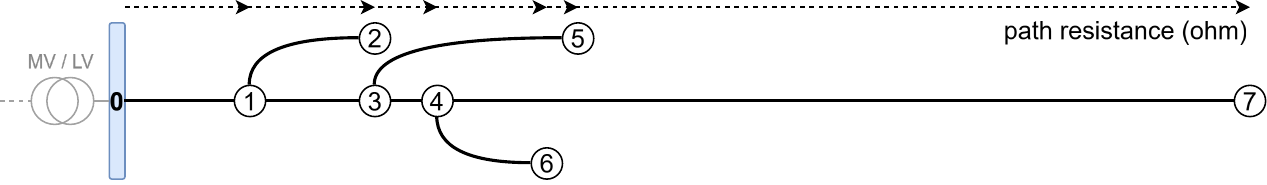}
	\caption{Diagram of the Belgian 8-node test feeder. Horizontal distance represent the path resistance from node~0.}
	\label{fig:dn_belg8n}
\end{figure}

\noindent Observe that all three approaches fully allocated the envelopes to node~1, as in this case node~2 has a significant locational penalty. When forcing allocation to node~2, by making node~1 as a non-participant, the envelope amount fell to roughly 50\% of the current values for every approach. It is also worth noting that \ac{LACE} and \ac{LP-DOE} results are precisely the same, showing the analytical understanding and computation of our proposed approach matches the conventional \ac{LP-DOE} available in the literature. Observe that in general \ac{NLP-DOE} presents smaller import and larger export (in absolute values) than the linear approaches due to the line losses (see Fig.~\ref{fig:nlp_thermal}).

\subsection{Realistic Test Feeder --- Belgium}

Fig.~\ref{fig:dn_belg8n} depicts the \ac{LV} feeder diagram. For each node, there are three customers and three participating flexible resources connected to it: a 10~kW solar \ac{PV}, a 11.5~kW bidirectional \ac{EV} charger and a 11.5~kW home battery. Hence the envelope bounds are $\underline{p}_m=-33$~kW and $\overline{p}_m=23$~kW. The transformer nominal rating is 65~kVA and we select a high-demand evening period in the Flobecq dataset to present a thermally-constrained case. The voltage-constrained case is demonstrated by replacing the existing transformer with a 200~kVA transformer.

\begin{figure}[!t]
	\centering
	\subfloat[]{%
		\includegraphics[width=1.00\linewidth]{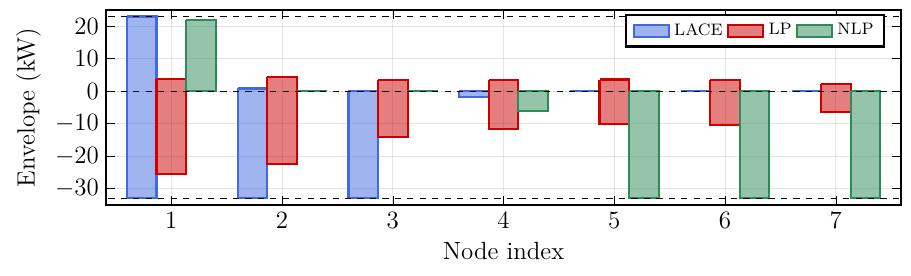}}
	\hfill
	\subfloat[]{%
		\includegraphics[width=1.00\linewidth]{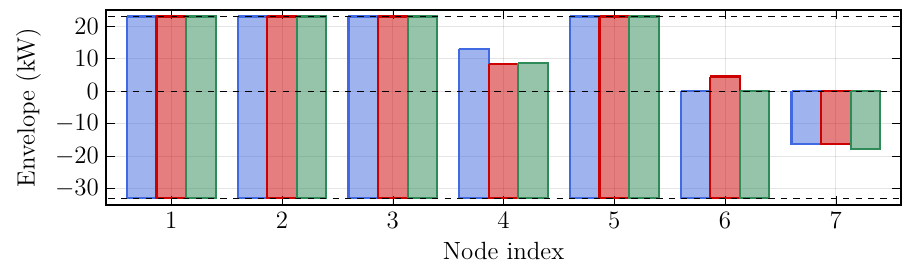}}
	\caption{DOE results for the Belgian test feeder: (a) thermally-constrained and (b) voltage-constrained case.}
	\label{fig:doe_belg}
\end{figure}

Fig.~\ref{fig:doe_belg} presents the \ac{DOE} calculation outcome for the (a) thermally- and (b) voltage-constrained case. In Fig.~\ref{fig:doe_belg}(a) the linear approaches (\ac{LACE} and \ac{LP}) resulted in different allocation, but the same combined amount of $-100.7$~kW export and 23.8~kW import, as there is no locational penalty is this case. In fact, the allocation for the \ac{LP-DOE} completely changes to an equally optimal solution when using a different solver. The \ac{NLP-DOE} approach fully allocated import envelope to node~1 and the export envelopes to nodes~4, 5, 6, 7 as they are further downstream (see Fig.~\ref{fig:dn_belg8n}).

Fig.~\ref{fig:doe_belg}(b) presents the voltage-constrained case. Observe that all \ac{DOE} approaches allocated more envelopes to nodes closer to the substation (1, 2, 3) and smaller envelopes to nodes further downstream (6, 7). The only exception being node~5 getting larger import \acp{DOE} than node~4, although node 4 is closer to the substation. To explain this, note that the voltage at the end of the main branch (node~7) is the only limiting factor (reaching 0.90~p.u.) and the common path between node~5 and 7 ($R_{33}=2(r_{01}+r_{13})$) is smaller than between node~4 and 7 ($R_{44}=2(r_{01}+r_{13}+r_{34})$). This results in $p_5^{solo} > p_4^{solo}$ and therefore an import allocation preference to node~5, rather than node~4. In summary, in the case of bifurcations, not only the electrical distance of the allocating node that matters, but rather the \textit{common electrical distance} (sensitivity) between the allocating node and the limiting factor node.

\subsection{Scalability}

We select representative parameters for a \ac{LV} test feeder and increment the number of nodes and line segments in steps of 10 to create more than 100 test feeders with different sizes, topologies and number of laterals. We run the \ac{DOE} calculation engines for every test feeder and analyze the computation efficiency.

\begin{figure}[!t]
	\centering
	\subfloat[]{%
		\includegraphics[width=0.50\linewidth]{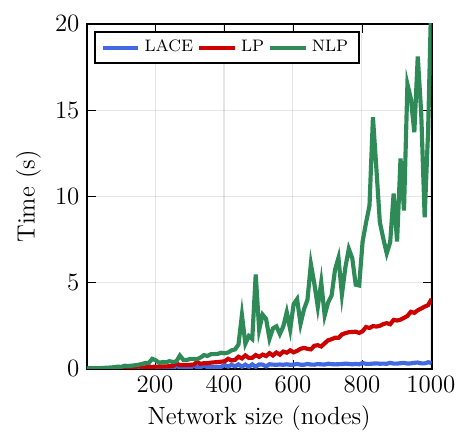}}
	\hfill
	\subfloat[]{%
		\includegraphics[width=0.48\linewidth]{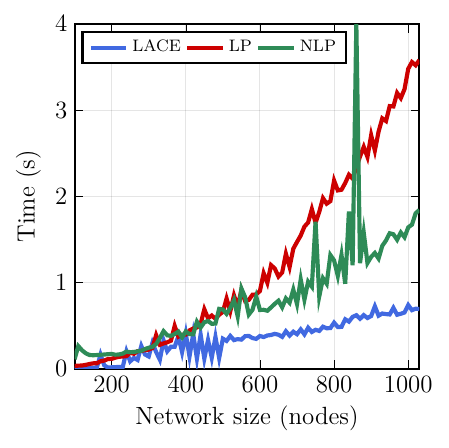}}
	\caption{\ac{DOE} engine computation time (import + export) for network with (a) chain topology and (b) tree topology with many laterals.}
	\label{fig:doe_scala}
\end{figure}

Fig.~\ref{fig:doe_scala} presents the wall-clock computation time for two edge cases: (a) radial networks with no laterals (chain topology); and (b) radial network with many laterals (tree topology). In general, all \ac{DOE} engines have provided efficient computation, with the longest computation time being under 20~s for the \ac{NLP-DOE} for a 1002-node network. Observe that the \ac{NLP-DOE} computation time considerably varies depending on the topology and number of nodes, presenting shorter computation time than the \ac{LP-DOE} for network topologies with many laterals.

The \ac{LACE} algorithm consistently presents the shortest calculation time. We envisage the \ac{LACE} algorithm will be a valuable tool to assist \acp{DSO} in computing \acp{DOE}, particularly as the \ac{DOE} problem formulation grows to multi-period and multi-phase with integers. \acp{DSO} can leverage on \ac{NLP} models for a more precise \ac{DOE} calculation, and leverage on linear approaches (\ac{LACE} and \ac{LP}) for more scalable approaches or as a warm start for the \ac{NLP-DOE}.

\section{Conclusion}

The paper presented two widely adopted optimization formulations with the linear and non-linear power flow models to calculate \acp{DOE} in distribution networks. Mathematical developments and numerical simulations unveiled how optimization engines compute and allocate the available network capacity to different locations on the network. Such insights can assist \acp{DSO}, aggregators, policy makers and academics alike to leverage \acp{DOE} and design different fairness policies. We also proposed a \ac{DOE} engine algorithm (LACE) that has shown a promising scalability trait which might be of interest for \acp{DSO} dealing with large-scale problems.

\section*{Acknowledgment}
This work is supported by FNR Luxembourg in the frame of the project DEFIE (INTER/FNRS/24/19145021/DEFIE).

\ifCLASSOPTIONcaptionsoff
  \newpage
\fi



\bibliographystyle{IEEEtran}
\bibliography{./bibtex/mybibfile}

\end{document}